\documentclass[prd,showpacs,amsmath,amssymb,superscriptaddress,nofootinbib]{revtex4-2}

\usepackage{graphicx}
\usepackage{dcolumn}
\usepackage{bm}
\usepackage{cases}
\usepackage{color}
\def\red{\color{red}}

\newcommand{\bea}{\begin{eqnarray}}
\newcommand{\eea}{\end{eqnarray}}
\newcommand{\pa}{\partial}
\newcommand{\nn}{\nonumber\\}

\newcommand{\hide}[1]{}

\begin{document}

\title{On the Cutoff Scale Identification of FLRW Cosmology in Asymptotically Safe Gravity}

\author{Chiang-Mei Chen} \email{cmchen@phy.ncu.edu.tw}
\affiliation{Department of Physics, National Central University, Zhongli, Taoyuan 320317, Taiwan}
\affiliation{Center for High Energy and High Field Physics (CHiP), National Central University, Zhongli,
Taoyuan 320317, Taiwan}


\author{Rituparna Mandal} \email{drimit.ritu@gmail.com} \thanks{corresponding author}
\affiliation{Department of Physics, National Central University, Zhongli, Taoyuan 320317, Taiwan}

\author{Nobuyoshi Ohta} \email{ohtan.gm@gmail.com}
\affiliation{Research Institute for Science and Technology, Kindai University, Higashi-Osaka, Osaka 577-8502, Japan}

\affiliation{Nambu Yoichiro Institute of Physics, Osaka Metropolitan University, Osaka 558-5585, Japan}

\date{\today}

\begin{abstract}
We examine Friedmann–Lema\^{\i}tre–Robertson–Walker cosmology, incorporating quantum gravitational corrections
through the functional renormalization group flow of the effective action for gravity. We solve the Einstein
equation with quantum improved coupling perturbatively including the case with non-vanishing classical cosmological
constant (CC) which was overlooked in the literatures. We discuss what is the suitable identification of the momentum
cutoff $k$ with time scale, and find that the choice of the Hubble parameter is suitable for vanishing CC
but not so for non-vanishing CC. We suggest suitable identification in this case.
The energy-scale dependent running coupling breaks the time translation symmetry and then introduces a new physical
scale.
\end{abstract}


\maketitle

\section{Introduction}
The effects of quantum gravity are crucial for establishing a consistent cosmological model spanning from
the early to late universe. Developing a consistent and predictive theory of quantum gravity to achieve this
is a significant challenge in theoretical physics. The perturbative approach in quantum gravity has been limited
due to the negative mass dimension of the Newton coupling. However, in recent years, the asymptotically safe
gravity has emerged as a promising scenario. The presence of a nontrivial fixed point renders the theory safe
from UV divergences~\cite{Reuter:1996cp, Souma:1999at}. The use of the functional renormalization group (FRG)
within the context of asymptotically safe gravity has provided new avenues to explore phenomena
in black hole physics and cosmology, extending from the UV scale to the IR
scale~\cite{Reuter:1996cp, Percacci:2017, Eichhorn:2019, Reuter:2019book}. This approach enables the examination
of consistent cosmological models across the entire UV to IR range.

Building on the framework of the FRG, the central element in the asymptotically safe gravity is the effective average
action, $\Gamma_{k}[g_{\mu \nu}]$. This action is constructed to describe gravitational phenomena at a momentum
scale $k$, accounting for the effects of quantum loops. To define the effective average action, a regulator
term $R_k$ is introduced, which suppresses the contributions from momentum modes $p$ below $k$. As a result, quantum
fluctuations below the FRG scale $k$ are excluded from the effective action, while modes with $p > k$ are fully
integrated out. More explicitly the FRG equation is \cite{Reuter:1996cp, Reuter:2019book, Wetterich:1993}
\bea
k \frac{d\Gamma_k}{dk} = \frac12 {\rm Tr} \left\{\left[ \Gamma_k^{(2)} + R_k \right]^{-1} k\frac{dR_k}{dk}\right\},
\label{frge}
\eea
where $\Gamma_k^{(2)}$ is the second variation of the effective average action.

However, solving fully the FRG equation in a realistic theory is challenging. To make progress, the Einstein-Hilbert
truncation is often employed. The action is defined as

\bea
\Gamma_k = \frac{1}{16 \pi G(k)} \int d^4 x \sqrt{-g}  (R(g) - 2 \Lambda(k)),
\eea
where $R(g)$ is the Ricci scalar for the Riemannian metric $g_{\mu\nu}$, $g$ is $\det(g_{\mu\nu})$,
$G(k)$ is the Newton coupling and $\Lambda(k)$ is the cosmological parameter.
The FRG ensures that the effective average action interpolates between the fundamental action,
devoid of quantum corrections in the UV regime, and the quantum effective action as $k \rightarrow 0$.
This interpolation makes the coupling ``constants'' scale dependent, and their evolution is governed by the FRG
equation~\cite{Reuter:1996cp, Bonanno:2001xi, Codello:2008vh}. For this reason, we call the Newton and
cosmological constants Newton coupling and cosmological parameter.
We show in the appendix~\ref{sec_SolRG} that the FRG equations yield scale-dependent expressions for these
couplings:
\begin{eqnarray}
G(k) &=& G_0 \left[ 1 - \omega G_0 k^2 + \omega_1 G_0^2 k^4 + \mathcal{O}\left( G_0^3 k^6 \right) \right],
\nonumber\\
\Lambda(k) &=& \Lambda_0 \left[ 1 - {\mu G_0 k^2} + \mu_1 G_0^2 k^4 + \mathcal{O}\left( G_0^3 k^6 \right) \right]
+ G_0 k^4 \left[ \nu + \nu_1 G_0 k^2 + \mathcal{O}(G_0^2 k^4) \right],
\label{eq_GLam}
\end{eqnarray}
where $G_0$ and $\Lambda_0$ are the Newton coupling and cosmological parameter at $k=0$. The dimensionless parameters
$\omega, \omega_1$ etc. depend explicitly on the choice of regularization scheme,
and we give the results for two schemes, the optimized and exponential cutoffs there.
However the results are qualitatively similar, and both $G(k)$ and $\Lambda(k)$ may be written compactly
to allow the $\Lambda_0 = 0$ solution by simply setting $\Lambda_0 = 0$ in~\eqref{eq_GLam}.
The solutions correspond to two distinct branches. The first branch, known as the Type IIa trajectory
or the separatrix, assumes $\Lambda_0 = 0$~\cite{Reuter:2001ag, Codello:2008vh,KO}.
It extends to $k \rightarrow 0$, smoothly approaching the Gaussian fixed point in the IR regime.
In contrast, the second branch, referred to as Type IIIa trajectories, corresponds to $\Lambda_0 \neq 0$.
This branch may not extend to $k \rightarrow 0$ due to the emergence of a singularity in the IR regime,
limiting its validity to a finite momentum scale. However, as discussed in \cite{Machado:2008}, the nonlocal
extension of the Einstein-Hilbert truncation can potentially resolve the infrared singularity in
the RG trajectories with a positive cosmological parameter.
It is crucial to note that the solutions for $\Lambda_0 = 0$ and $\Lambda_0 \neq 0$ are derived separately,
and we will see that these give distinct quantum corrections for the two branches.
Nevertheless, the coefficients may be written uniformly like~\eqref{eq_GLam}.
We note that the quantum correction to the cosmological parameter begins at the $k^4$ order for $\Lambda_0 = 0$,
while it starts at the $k^2$-order for $\Lambda_0 \neq 0$, which differs from previous reports in
the literature~\cite{Reuter:2001ag, MRHW_2018}. This distinction shows the need to consider quantum-improved
cosmological solutions for the two cases separately.

Quantum effects can be incorporated into cosmological studies at various levels. In this work, we incorporate
quantum corrections into the Einstein equations by replacing the ordinary Newton coupling and cosmological
parameter with their scale-dependent counterparts. This approach allows us to explore the impact of quantum gravity
on cosmology. A crucial step in this setup is the identification of the IR momentum scale $k$ with a suitable
physical cutoff. Unlike black hole physics, where the cutoff scale can often be determined from physical mechanisms,
cosmology lacks a direct method for such identification. Existing literature addresses this challenge by
expressing $k$ in terms of physical quantities such as particle momenta or the spacetime curvature.
Most natural and common choice in cosmology is to identify $k$ with the Hubble parameter, which is an almost unique
physical scale. Various studies have explored Friedmann–Lema\^{\i}tre–Robertson–Walker (FLRW) cosmology
choosing different cutoff scales~\cite{Bonnano:2002plb, BONNANO20041, Bentivegna_2004, Reuter:2005jcap,
Bonanno:2007wg, Bonanno274, BF_2017, Moti:2019, lit1, Platania_2020}. In this work, we aim to determine a suitable
cutoff scale motivated by the classical solutions at late times. In general, different choices for the cutoff
scale can lead to varying cosmic evolution scenarios. This raises a critical question: how can we identify
which cutoff choices are more viable? To address this, we investigate the quantum-corrected late-time
behaviour of the Hubble parameter and the scale factor. Since quantum corrections are expected to be small
at late times, the quantum-corrected solutions should closely resemble the classical solutions with
minor quantum improvements. This implies that the perturbations introduced by quantum corrections must
not be so large as to significantly alter the form of the solutions at late times.
Through this analysis, we can identify a way to assess the viability of different cutoff scale choices.

We study the FLRW universe using the FRG-improved Einstein equations. This approach leads to a set
of differential equations involving the Hubble parameter, energy density, the scale-dependent Newton coupling,
the cosmological parameter, and a cutoff function. An important input for defining these equations is
the requirement that the left-hand side of the Einstein equations is covariantly conserved, resulting in
a modified continuity equation. This equation establishes a relationship between the matter density,
the time-dependent Newton coupling, and the cosmological parameter. Many cosmological studies have employed
this modified continuity equation approach~\cite{Bonanno:2007wg, Reuter:2003ca, Reuter:2004nv, lit1,
 Mandal:2020umo}. It is important to note that this method differs from the consistency approach,
which enforces the covariant conservation of both the energy-momentum tensor and the left-hand side of
the Einstein equations~\cite{Bonanno:2001xi, Bonnano:2002plb, Mandal:2019xlg}. In contrast, the modified continuity
equation approach focuses on the conservation of the left-hand side of the Einstein equations only.
Using this framework, we investigate the cosmological evolution for different cutoff choices across
two trajectory branches: one with $\Lambda_0 = 0$ and the other with $\Lambda_0 \neq 0$.
For the $\Lambda_0 = 0$ branch, we consider cutoff identifications where $k$ is either the classical Hubble parameter
or a function of the quantum Hubble parameter, and find that the choice of the classical Hubble parameter gives
reasonable cosmology, and another choice of quantum Hubble parameter gets some constraint.
For the $\Lambda_0 \neq 0$ branch,
the Hubble parameter becomes constant in late time, so it is not suitable to describe the time development.
We find that the choice of the inverse cosmic time successful for $\Lambda_0=0$ is not suitable.
Instead, we consider a time-dependent cutoff scale proportional to the inverse of the classical scale factor,
$k \propto 1/a_\mathrm{cl}(t)$, which is suggested by the scalar curvature for compact space.
We find that this choice ensure consistent solutions with the classical Hubble parameter and scale factor.

This paper is organized as follows. In sect.~\ref{imp_Fie_eq}, we present the improved field equations within
the FRG framework. Section~\ref{FLRW_cosmo_lambda0} focuses on FLRW cosmology with $\Lambda_0 = 0$.
We first summarize the classical solution for $\Lambda_0 = 0$ in subsect.~\ref{classical_lamba0}.
In subsect.~\ref{lambda0chu}, we use the identification of the momentum scale with the classical Hubble parameter
$H_{\rm cl}$ and find reasonable behavior of the late-time cosmology with quantum corrections.
In subsect.~\ref{lambda0qhub}, we modify the identification to use the quantum Hubble parameter, discuss
how the solution is different.
In sect.~\ref{lambane0}, we consider FLRW cosmology with $\Lambda_0 \neq 0$.
We then summarize the classical solution in subsect.~\ref{classical_lambdanon0}.
In this case, we show in subsect.~\ref{lambdanon0hub} that the same choice of the identification as
the $\Lambda_0=0$ case, namely the inverse of the cosmic time does not give good late-time behavior, which
does not reproduce the classical behavior in the classical limit. So it is not a viable identification.
Then in subsect.~\ref{lambdanon0scale}, we show that good identification is obtained by using the scale factor,
ensuring consistent solutions at late times.
Finally, in sect.~\ref{conclu}, we summarize our results.

\section{The Improved Field Equation}
\label{imp_Fie_eq}

In this section, the FRG improvement is applied at the ``equation level'' by replacing the ordinary Newton
coupling and cosmological constant with the scale-dependent running coupling parameters $G(k)$
and $\Lambda(k)$ in the Einstein equation. We take the viewpoint that these running coupling parameters capture
quantum effects at the leading order. The quantum improved Einstein equation can thus be written as
\begin{eqnarray}
R_{\mu\nu} - \frac{1}{2} R g_{\mu\nu} = - \Lambda(t) g_{\mu\nu} + 8 \pi G(t) T_{\mu\nu}~.
\label{IEE}
\end{eqnarray}
In writing down this equation, $G(k)$ and $\Lambda(k)$ are replaced with $G(t)$ and $\Lambda(t)$ to facilitate
the study of cosmology, where the infrared cutoff parameter becomes a function of the cosmic time $t$,
so that $k=k(t)$.

We consider a spatially flat, homogeneous, and isotropic cosmological model described by
the FLRW spacetime:
\begin{eqnarray}
ds^2 = - dt^2 + a^2(t) (dx^2 + dy^2 + dz^2),
\label{FLRW}
\end{eqnarray}
where $a(t)$ is the scale factor. The associated geometric quantities for this spacetime are given by
\begin{equation}
R_{tt} = -3 \frac{\ddot a}{a}, \qquad R_{ii} = a \ddot a + 2 \dot a^2, \qquad
R = 6 \left( \frac{\ddot a}{a} + \frac{\dot a^2}{a^2} \right)~.
\label{GQ}
\end{equation}
Moreover, The cosmic matter is assumed to be a perfect fluid, described by the energy-momentum tensor
\begin{align}
T^\mu{}_\nu = \mathrm{diag}( - \rho, p, p, p ),
\label{EMT}
\end{align}
where $\rho$ is the energy density, $p$ is the pressure and they are related by the equation of state
\bea
p = w \rho.
\label{eos}
\eea
From these, the first modified Friedmann equation can be written as
\begin{align}
H^{2} = \frac{{\dot{a}}^{2}}{a^2} &= \frac{8 \pi}{3} G(t) \rho + \frac{\Lambda(t)}{3} \,~.
\label{FFE}
\end{align}
In addition, the left-hand side of the Einstein equation is covariantly conserved, expressed as
$D^\mu ( R_{\mu\nu} - R g_{\mu\nu}/2 ) = 0$, which is a mathematical identity in Riemannian geometry
known as the Bianchi identity. Consequently, the entire right-hand side of the Einstein equation must also
be covariantly conserved $D^\mu \left( - \Lambda g_{\mu\nu} + 8 \pi G T_{\mu\nu} \right) = 0$.
This leads to the continuity equation:
\begin{align}
\dot{\rho} + 3 H (p + \rho) &= - \frac{8 \pi \rho \dot{G} + \dot{\Lambda}}{8 \pi G(t)}\,.
\label{modifiedcon}
\end{align}
Then there are two possible standpoints at this stage.
\vspace{2mm}

\noindent
{\bf (i) Consistency condition}

In the absence of quantum effects, in particular if the Newton coupling $G$ and cosmological constant $\Lambda$ are
independent of time, the energy-momentum tensor is conserved $D^\mu T_{\mu\nu} = 0$. So it may appear natural
to require this.
This leads to the ordinary continuity equation, and both the LHS and RHS of~\eqref{modifiedcon} should
vanish~\cite{Bonanno:2001xi},
\begin{align}
\dot{\rho} + 3 H (p + \rho) = 0, \qquad
8 \pi \rho \dot{G} + \dot{\Lambda} = 0.
\label{consistency}
\end{align}
However in this case, the second equation gives
\bea
\rho = - \frac{\dot\Lambda(t)}{8\pi \dot G(t)}.
\eea
As we will discuss later, if we identify the momentum scale $k$ with inverse of the cosmic time $t$,
this is translated into
\bea
\rho = - \frac{\pa_k \Lambda(k)}{8 \pi \pa_k G(k)}.
\eea
We see from the low energy expansions in Eq.~\eqref{eq_GLam}, the
leading term of $\pa_k G(k)$
is negative (for $\omega > 0$). So the sign of the energy density $\rho$
depends entirely on the sign of
the coefficient of the leading quantum correction to the cosmological
parameter.
For $\Lambda_0 \ne 0$, it is negative, giving negative $\rho$.
For $\Lambda_0 = 0$, it is positive (for $\nu > 0$), resulting in
positive $\rho$.
The case of $\Lambda_{0}=0$ is often considered, and discussed in detail
especially in Ref.~\cite{Bonanno:2001xi}.
However, in our case, there is another possibility, which we employ to
identify the cutoff scale in both cases.

\vspace{2mm}

\noindent
{\bf (ii) Modified continuity equation}

When we consider the quantum effects, the Newton coupling and cosmological parameter depend on the cosmic time. This arises because of the quantum effects of gravity. It is then expected that they also contribute to the
energy-momentum. It must be $G$ in front of the energy-momentum tensor and $\Lambda$ that give such contribution
in the present formulation. This suggests that we should include the time derivative of $G(t)$ and $\Lambda(t)$
in Eq.~\eqref{modifiedcon} as the energy-momentum from the gravity, instead of making its both sides separately
vanish.
This modified equation has been employed in various studies of cosmology within the asymptotic safety
framework~\cite{Bonanno:2007wg, Reuter:2003ca, Reuter:2003ca, Reuter:2004nv, lit1, Mandal:2020umo}.
This is weaker condition than (i) and gives interesting cosmology for $\Lambda_0 \neq 0$.
In this paper, we take this condition.

Let us write the modified continuity equation as
\begin{align}
8\pi \partial_{t}\left[G(t)\rho +\frac{\Lambda(t)}{8\pi}\right]=-24\pi (1+w)HG(t)\rho \,.
\label{ce}
\end{align}
where the equation of state~\eqref{eos} has been used.
Substituting $G(t)\rho$ from Eq.~\eqref{FFE} into the above equation, we obtain
\begin{align}
\dot{H}=-\frac{1}{2}(3+3w)\left [ H^{2}-\frac{1}{3}\Lambda(t) \right ]
= - \alpha \left( H^2 - \frac{\Lambda (t)}3 \right)\,,
\label{eq_Ein_H}
\end{align}
where $\alpha \equiv 3 (1 + w)/2$.
Furthermore, the energy density $\rho(t)$ can be expressed in terms of the solution for $H(t)$ and
the cosmological parameter $\Lambda(t)$:
\begin{eqnarray}
\rho = \frac3{8 \pi G(t)} \left( H^2 - \frac{\Lambda(t)}3 \right). \label{eq_Ein_rho}
\end{eqnarray}
We shall solve the system of differential equations given by~\eqref{eq_Ein_H} and~\eqref{eq_Ein_rho}
for $H(t)$ and $\rho(t)$, under the condition that $H(t) \neq 0$ for the prescribed form of $G(t)$ and $\Lambda(t)$.
It is important to note that to solve Eq.~\eqref{eq_Ein_H} for the Hubble parameter $H(t)$, only the form of
the external function $\Lambda(t)$ is required. The running Newton coupling $G(t)$ only appears
in Eq.~\eqref{eq_Ein_rho} when solving for $\rho(t)$ in terms of $H(t)$ and $\Lambda(t)$.
To determine $\Lambda(t)$, we have to make identification of the infrared momentum scale $k$ with the physical
cutoff scale. Existing literature often expresses $k$ in terms of scales, such as particle momenta, field strengths,
or the curvature of spacetime~\cite{Bonanno:2001xi}. However, there is no satisfactory principle to directly
correlate the FRG scale $k$ with the physical scale in cosmology, similar to the consistent thermodynamics
for quantum improved black holes~\cite{Chen:2022xjk, Chen:2023wdg}. For the FLRW universe,
homogeneity and isotropy imply that the only physical scale is the Hubble parameter which is determined
by the cosmic time {\red $t$}. Thus we have
\begin{eqnarray}
G(t)\equiv G(k=k(t)),\qquad \Lambda(t) \equiv \Lambda(k=k(t))~.
\label{gk}
\end{eqnarray}
A natural choice of the cutoff identification is in terms of the Hubble parameter $H(t)$.
We still have the possibility of using classical or quantum Hubble parameters. In the next section,
we first consider FLRW cosmology for $\Lambda_0=0$ using the identification with the classical Hubble parameter,
and then the quantum Hubble. In the latter case, we consider a general form of
$k(H) = \xi G_0^{(\beta - 1)/2} H^\beta$, where $\xi$ and $\beta$ are constants.

\section{FLRW Cosmology for $\Lambda_0=0$}
\label{FLRW_cosmo_lambda0}

In this section, we discuss the classical and quantum-corrected late-time solutions for $H(t)$, which, in turn,
provide the solution for the scale factor $a(t)$ for $\Lambda_0=0$.
This corresponds to a unique trajectory that reaches the Gaussian fixed point as $k \rightarrow 0$,
with $\Lambda_{0}=0$; this is commonly referred to in the literature as the Type IIa or separatrix
trajectory~\cite{Reuter:2001ag, Codello:2008vh}.
The second branch is discussed in sect.~\ref{lambane0}.
%
\subsection{Classical solution}
\label{classical_lamba0}
The classical solution for the Hubble parameter, denoted as $H_\mathrm{cl}(t)$, in a single component
universe specifically for radiation $(w = 1/3)$ and matter ($w = 0$) can be obtained from
Eq.~\eqref{eq_Ein_H} by setting $G(t) = G_{0}$ and $\Lambda(t) = \Lambda_0$.
The classical solution for the Hubble parameter with $\Lambda_0 = 0$ is given by
\begin{equation}
H_\mathrm{cl}(t) = \frac{1}{\alpha (t - c')},
\label{clHlam0}
\end{equation}
where $c'$ is an integration constant. Because the system has the time translation symmetry in Einstein equation,
the free time parameter $c'$ does not carry physical significance and we can set $c' = 0$ or absorb it into
the definition of the cosmic time $t$:
\begin{equation}
H_{\mathrm{cl}}(t)=\frac{1}{\alpha t}
\label{simhlamo}
\end{equation}
By this, we fix the time origin in the system.
The classical scale factor $a_\mathrm{cl}(t)$ is then derived by integrating the Hubble parameter as
\begin{equation}
a_\mathrm{cl}(t) =  a_0\, t^{1/\alpha}~.
\label{clalam0}
\end{equation}
The energy density can then be expressed using the classical Hubble parameter solution from Eq. \eqref{clHlam0}
as follows:
\begin{equation}
\rho_\mathrm{cl}(t) =  (3/8 \pi G_0) /(\alpha t)^2 .
\label{clrho0}
\end{equation}


\subsection{The identification with the classical Hubble parameter}
\label{lambda0chu}

We begin by discussing the quantum-corrected solutions and their implications. The unique physical scale
in this case is the Hubble parameter. Here we consider the classical Hubble parameter in the identification.
So we take
\bea
k = \xi' H_{\rm cl}(t) = \frac{\xi}{t} \, \propto a_{\rm cl}^{-\alpha}(t),
\label{id0}
\eea
where $\xi'$ and $\xi(=\xi'/\alpha)$ are dimensionless parameters of order 1.
By substituting this cutoff scale into Eq.~\eqref{eq_GLam},
the running gravitation coupling and the cosmological coupling can be written as a function of time:
\begin{eqnarray}
\label{eq_GLam0_t}
G(t) &=& G_0 \left[ 1 - \tilde\omega G_0 t^{-2} + \tilde\omega_1 G_0^2 t^{-4}
 + \mathcal{O}\left( G_0^3 t^{-6} \right) \right],
\nonumber\\
\Lambda(t) &=& t^{-2} \left[ \tilde\nu G_0 t^{-2} + \tilde\nu_1 G_0^2 t^{-4}
 + \mathcal{O}\left( G_0^3 t^{-6} \right) \right],
\end{eqnarray}
where $\tilde\omega \equiv \omega \xi^2, \tilde\omega_1 \equiv \omega_1 \xi^4$ and
$\tilde\nu \equiv \nu \xi^4, \tilde\nu_1 = \nu_1 \xi^6$. Note that the time translation symmetry is
no longer there by these parameters.

Substituting the cosmological parameter~\eqref{eq_GLam0_t} into~\eqref{eq_Ein_H}, and taking the ansatz for $H(t)$ as
\begin{equation}
H(t) = \frac{1}{\alpha t} \left( 1 + \frac{c}{t} + \frac{c_1}{t^2} + \frac{c_2}{t^3} + \frac{c_3}{t^4}
+ \cdots \right)~,
\label{eq_ansa}
\end{equation}
we can determine the the above constants by comparing terms order by order in $1/t$.
We then obtain
\begin{equation}
H(t) = \frac1{\alpha t} \left[ \left( 1 + \frac{c}{t} + \frac{c^2}{t^2} + \frac{c^3}{t^3}
+ \frac{c^4}{t^4} + \cdots \right) - \frac{\alpha^2 \tilde\nu G_0}{3 t^2} \left( 1 + \frac{c}{t}
+ \frac{c^2}{t^2} + \frac{c^2}{3 t^2} - \frac{\alpha^2 \tilde\nu G_0}{9 t^2} + \cdots \right)
- \frac{\alpha^2 \tilde\nu_1 G_0^2}{9 t^4} \left( 1 + \cdots \right) \right],
\label{qch}
\end{equation}
where $c$ is a new constant parameter.
The appearance of one undetermined parameter $c$ is expected, as Eq.~\eqref{eq_Ein_H} is a first-order
differential equation, which naturally provides one integration constant.
We see the terms in the first bracket sums up to $1/(t-c)$, quite similar to the classical solution.
However we cannot remove this parameter by absorbing it to the cosmic time $t$ because it changes the form of
the cosmological parameter term, in contrast to classical case. But in the absence of the quantum corrections
($\tilde\nu=\tilde\nu_1=0$), $c$ becomes identical with the classical parameter.

We now show that this new parameter $c$ has physical
meaning. This point was not discussed in~\cite{Mandal:2020umo}.
The solution, up to fourth order in $1/t$, can be re-expressed  as
\begin{equation}
H(t) = \frac1{\alpha t} \left[ 1 - \frac{\alpha^2 \tilde\nu G_0}{3 t_c t}\left( \frac{t_c}{t} - 1 \right)
- \left( \frac{\alpha^2 \tilde\nu G_0}{3 t_c t} \right)^2\left( \frac{t_c}{t} - 1 \right)
- \left( \frac{\alpha^2 \tilde\nu G_0}{3 t_c t} \right)^3 \left( \frac{t_c}{t} - 1 \right)
+ \mathcal{O}\left( t^{-4} \right) \right],
\label{newqch}
\end{equation}
where the critical time scale $t_c$ is defined by
\begin{equation}
t_c = \alpha^2 \tilde\nu G_0/3 c~.
\end{equation}
The quantum parameter $\tilde\nu$, which defines the critical time, is always positive for both
the optimized and exponential cutoffs: specifically, including the dimensionless parameter $\xi$,
$\tilde\nu = \xi^4/8\pi$ for the former case,
as defined in Eq.~\eqref{eq_Optimized_0} and $\tilde\nu = \xi^4 \zeta(3)/2\pi$ for the latter
in Eq.~\eqref{eq_exp_0}. Therefore, this positive critical time scale indicates that the leading quantum
correction has an attractive effect (AdS-phase) for which the expansion rate of the FLRW universe slows
down compared with the speed of classical expansion, $H(t) < H_\mathrm{cl}(t) = 1/\alpha t$, for $t < t_c$.
Conversely, it becomes repulsive (dS-phase) and makes the expansion rate bigger than the classical expansion rate
when $t > t_c$. The factor $(t_c/t - 1)$ is not universal, but appears up to $t^{-3}$-order and
the remaining terms are much smaller.
So the critical time scale, which distinguishes the AdS and dS phases, is qualitatively correct.

The value of the parameter $c$ can be estimated by physical observation data. We can get approximated idea about
the value of $c$ is by setting the Hubble parameter $H(t_{0})=H_0$ at present time $t_0$, with $\alpha = 3/2$
($w= 0$ matter dominated era) in the solution $H(t_0)=1/(\alpha(t_0-c))$ with the leading correction
from Eq.~\eqref{qch}. From the condition that this should be the current time Hubble parameter $H_0$ at $t_0$, we get
\begin{equation}
\label{eq_c}
c = t_0 - \frac1{\alpha H_0}, \qquad t_0 = (1.373 \pm 0.012) \times 10^{10} \;
\mathrm{years}, \quad
1/H_0 = 1.45 \times 10^{10} \; \mathrm{years}.
\end{equation}
This means that in general $c$ is a huge positive number of the same order or one order smaller than $t_0$,
and thus the value of $t_c$ is very small. This implies the quantum correction always provides repulsive effect
in late time of FLRW cosmology $t \gg t_c$. Note that the late time solution~\eqref{qch} is valid for $t > c$,
implying~\eqref{eq_c} is a consistent condition.
The quantum corrected scale factor $a(t)$ can be calculated integrating the Hubble parameter
$H(t)=\dot{a}(t)/a(t)$ as follows:
\begin{align}
a(t) &= a_0  t^{1/\alpha} \left[ 1 - \frac{c}{\alpha t} - \frac{(\alpha - 1) c^2}{2 \alpha^2 t^2}
 + \frac{\alpha \tilde\nu G_0}{6 t^2} + \mathcal{O}\left( t^{-3} \right) \right] \nonumber \\
& = a_0  t^{1/\alpha} \left[ 1 - \frac{\alpha \tilde\nu G_0}{3 t_c t} - \frac{\alpha^2 (\alpha - 1)
 \tilde\nu^2 G_0^2}{18 t_c^2 t^2} + \frac{\alpha \tilde\nu G_0}{6 t^2} + \mathcal{O}\left( t^{-3} \right) \right]~.
\label{qscale}
\end{align}
Substituting $\Lambda(t)$ from Eq.~\eqref{eq_GLam0_t} and the Hubble parameter $H(t)$ from Eq.~\eqref{qch}
in Eq.~\eqref{eq_Ein_rho}, we obtain the quantum corrected energy density to be
\begin{equation}
\rho(t) = \frac{3}{8 \pi \alpha^2 G_0 t^2} \left[ 1 + \frac{2 c}{t} + \frac{3 c^2}{t^2}
 + \frac{\tilde\omega G_0 - \alpha^2 \tilde\nu G_0}{t^2}  + \mathcal{O}\left( t^{-3} \right) \right]~.
\label{qeneden}
\end{equation}
Note that, by setting $\tilde{\nu} = \tilde{\omega}=0$ and $c = 0$, we recover the classical result of energy density
in FLRW cosmology. Here we can have a further physical understanding for the existence of the critical time $t_c$.
The leading-order quantum correction to the cosmological parameter, $\Lambda \simeq \tilde\nu G_0/t^4 > 0$,
introduces a repulsive force. However, the integration constant $c$ increases the energy density $\rho(t)$
which provides an attractive force. The entire effect is the competition of these two opposite contributions,
with the critical time $t_c$ characterizing the balance point between them, as can be seen in Eq.~\eqref{newqch}.

\subsection{The identification by the Hubble parameter with quantum correction}
\label{lambda0qhub}

Here we consider some general type of cutoff identification as mentioned before, defined as a function of
the quantum Hubble parameter $H(t)$ in the form
\bea
k(H) = \xi G_0^{(\beta - 1)/2} H^\beta,
\eea
where $\xi$ is a dimensionless constant. To ensure that $\xi$ in the cutoff scale is dimensionless, unlike in
\cite{Mandal:2020umo}, we introduce  $G_{0}$ within the cutoff scale.
Three specific values of $\beta = 1/4, 3/4, 1$ were considered in~\cite{Mandal:2020umo}, and we study which of
these values are viable. By substituting this cutoff choice into~\eqref{eq_GLam} with $\Lambda_0=0$,
we obtain the leading correction to the cosmological parameter as
\begin{equation}
\Lambda = 3 \gamma H^{4 \beta} + \cdots,
\label{lamkh}
\end{equation}
where $\gamma = \xi^4 \nu G_0^{2\beta - 1}/3$. Putting \eqref{lamkh} in~\eqref{eq_Ein_H}, the Einstein
equation~\eqref{eq_Ein_H} for $H(t)$ becomes
\begin{equation}
\frac{dH}{dt} = - \alpha ( H^2 - \gamma H^{4 \beta} )~.
\label{einh}
\end{equation}
Using the relation $H^{-1} (d H/d t) = a (d H/d a)$, we can rewrite the above equation in the form
\begin{equation}
- \frac{H^{2 - 4 \beta}}{ H \left( \gamma - H^{2 - 4 \beta} \right)} d H = - \alpha \frac{da}{a}~.
\label{neweinh}
\end{equation}
Integrating the above equation, we obtain the solution of $H(t)$ in terms of scale factor $a(t)$ as
\begin{equation}
H = a^{-\alpha} \left( \gamma a^{-\alpha (4 \beta - 2)} - C_0 \right)^{-1/(4 \beta - 2)}.
\label{solein}
\end{equation}
where $C_{0}$ is the integration constant. To determine this, we require the scale factor
$a(t_0) = 1$ and $ H(t_0) = H_0$ at present time, yielding
\begin{equation}
C_0 = \gamma - H_0^{2 - 4 \beta}~.
\label{intc0}
\end{equation}
By substituting the integration constant $C_0$ into Eq.~\eqref{solein}, we get the final form
of the quantum corrected Hubble parameter $H(t)$:
\begin{equation}
H = H_0 a^{-\alpha} \left[ 1 + \gamma H_0^{4 \beta - 2} \left( a^{-\alpha (4 \beta - 2)} - 1 \right)
 \right]^{-1/(4 \beta - 2)}~.
\label{finasolh}
\end{equation}
We will examine the asymptotic behaviour of the Hubble parameter at $t \to \infty$. Our aim is to determine if there
exists a critical value of the cutoff scale exponent $\beta$ at which the entropy transitions from constant to
divergent occurs at $t \to \infty$, as previously discussed~\cite{Mandal:2020umo}. In the limit of infinitely large
time, the scale factor $a$ becomes $\infty$ for spatially flat FLRW universe. We then observe from Eq.~\eqref{finasolh}
two distinct behaviours of $H(t)$ depending on the value of $\beta$:
\begin{equation}
H \approx \left\{
\begin{array}{cl} {\tilde H}_0 a^{-\alpha}, & \quad \beta > 1/2, \\
\tilde\gamma \ \ , & \quad \beta < 1/2, \end{array} \right.
\end{equation}
where ${\tilde H}_0 = H_0 (1 - \gamma H_0^{4 \beta - 2})^{-1/(4 \beta - 2)}$ and
$\tilde\gamma = \gamma^{-1/(4 \beta - 2)}$. Furthermore, the scale factor $a(t)$ takes the form
\begin{equation}
a(t) \approx \left\{
\begin{array}{cl} (\alpha {\tilde H}_0)^{1/\alpha} t^{1/\alpha}, & \quad \beta > 1/2, \\
\mathrm{e}^{\tilde\gamma\, t}, & \quad \beta < 1/2, \end{array} \right..
\label{asympa}
\end{equation}
From the asymptotic behaviour of $H(t)$ and the scale factor $a(t)$, it is clear that $\beta = 1/2$ is a critical
value which changes the scale factor behaviour from a power law ($\beta > 1/2$) to exponential ($\beta < 1/2$).

In order to find the higher order terms in the solution of the Hubble parameter $H(t)$  at late time,
we expand $H(t)$ as
\begin{equation}
H(t) = \left\{
\begin{array}{cl}
{\tilde H}_0 a^{-\alpha} \left[ 1 + \Delta a^{-\alpha (4 \beta - 2)} \right]^{-1/(4 \beta - 2)}
 = {\tilde H}_0 a^{-\alpha} \left[ 1 - \frac{\Delta}{4 \beta - 2} a^{-\alpha (4 \beta - 2)} + \cdots \right],
 & \quad \beta > 1/2 \\
\tilde\gamma \left[ 1 + \Delta^{-1} a^{- \alpha (2 - 4 \beta)} \right]^{-1/(4 \beta - 2)}
 = \tilde\gamma \left[ 1 - \frac1{\Delta (4 \beta - 2)} a^{- \alpha (2 - 4 \beta)} + \cdots \right],
 & \quad \beta < 1/2 \end{array} \right.,
\label{solha}
\end{equation}
where $\Delta = (\tilde H_0/H_0)^{4 \beta - 2} - 1$. Integrating the above equation, we obtain the late time
behaviour of the scale factor $a(t)$ with leading order correction as
\begin{equation}
a(t) = \left\{
\begin{array}{cl}
(\alpha {\tilde H}_0)^{1/\alpha} t^{1/\alpha} \left[ 1 + \frac{\Delta (4 \beta - 1)}{\alpha (4 \beta - 2)^2}
 (\alpha \tilde H_0)^{- (4 \beta - 2)} t^{-(4 \beta - 2)} + \cdots \right], & \quad \beta > 1/2 \\
\exp\left[ \tilde\gamma t - \frac1{\Delta \alpha (2 - 4 \beta)^2} \mathrm{e}^{- \alpha \tilde\gamma (2 - 4 \beta) t}
 + \cdots \right], & \quad \beta < 1/2 \end{array} \right..
\label{solasc}
\end{equation}
The exponential solution for the cutoff choice with $\beta < 1/2$ suggests that this choice is not viable.
For $\beta < 1/2$, large momentum scales $k$ cause significant perturbations, leading to exponential
expansion which is completely different from the classical solution. On the other hand, for $\beta > 1/2$,
the behaviour follows a power law similar to the classical case,
as the perturbations remain small enough to maintain this expected behaviour with the small quantum corrections.
Thus, the cutoff identification is only valid for $\beta > 1/2$ for late time.

For the sake of completeness, let us examine the cosmic evolution when the exponent of the momentum cutoff scale
is exactly $\beta = 1/2$. This gives $k = \xi G_{0}^{-1/4} H^{1/2} = \xi \sqrt{m_\mathrm{Pl}} H^{1/2}$
and the leading order cosmological parameter becomes $\Lambda = \xi^4 \nu H^{2}$.
The Friedmann equation then takes the form
\begin{equation}
\frac{d H}{d t} = - \alpha \left( 1 - \frac{\xi^4 \nu}{3} \right) H^2~.
\label{fridb12}
\end{equation}
Solving this equation, we find the Hubble parameter as a function of cosmic time:
\begin{equation}
H(t) = \frac{1}{\alpha (1 - \xi^4 \nu/3) (t - c)},
\end{equation}
where $c$ is an arbitrary integration constant. Integrating $H(t)$, the scale factor can be written as
\begin{equation}
a(t) = a_0 (t - c)^{\frac{1}{\alpha (1 - \xi^4 \nu/3)}}~.
\end{equation}
We observe that the solution maintains a power-law behaviour, similar to the classical case, unlike the solutions
for $\beta < 1/2$. Although the cutoff scale $k$ is large, scaling with the Planck mass, the running
cosmological parameter does not depend on the Planck mass and remains small.
Thus, $\beta \geq 1/2$ is a viable cutoff choice and $\beta = 1/2$ represents the critical value,
below which $\beta$ are not good choices for late time.

\section{FLRW cosmology for $\Lambda_0 > 0$}
\label{lambane0}

In this section, we begin by summarizing the classical solutions for $H(t)$, $a(t)$ and $\rho(t)$
for the branch of trajectories where $\Lambda = \Lambda_0 >0$.

\subsection{Classical solution}
\label{classical_lambdanon0}

To calculate the classical solution
for the Hubble parameter $H_{\mathrm{cl}}(t)$ in a universe dominated by a cosmological parameter,
the classical Einstein equation~\eqref{eq_Ein_H} with $\Lambda = \Lambda_0 > 0$ yields:
\begin{equation}
H_{\mathrm{cl}}(t)=\sqrt{\Lambda_0/3} \, \frac{1 + \bar{c}\, \exp\left(-2\alpha\sqrt{\Lambda_0/3} \,t\right)}
{1-\bar{c} \, \exp\left(-2\alpha\sqrt{\Lambda_0/3} \,t\right)},
\label{hlamne0}
\end{equation}
where $\bar{c}$ is the integration constant.
This exact solution has two branches, depending on the sign of the free parameter $\bar{c}$.
Again using the time translation symmetry to shift the cosmic time $t$, we can set $|\bar c|=1$.
Depending on the sign of $\bar{c}$, the classical solution then becomes
\begin{equation}
H_\mathrm{cl}(t)= \begin{cases} \sqrt{\Lambda_0/3} \,\tanh\left( \alpha \sqrt{\Lambda_0/3}\, t \right),
& \mathrm{sign}(\bar{c}) = -1, \\ \sqrt{\Lambda_0/3} \, \coth\left( \alpha \sqrt{\Lambda_0/3} \,t \right),
& \mathrm{sign}(\bar{c}) = 1 . \end{cases}
\label{solhne0}
\end{equation}
At late times, both branches asymptotically approach a constant value of  $\sqrt{\Lambda_0/3}$,
consistent with the expected behaviour of a cosmological constant dominated universe. However, we select
the branch with $\mathrm{sign}(\bar{c}) = -1$, as it provides a continuous solution at $t=0$,
unlike the $\mathrm{sign}(\bar{c}) = 1$ branch, which has a discontinuity at $t=0$.
For $\mathrm{sign}(\bar{c}) = -1$, the exact late-time behaviour is written as
\begin{equation}
H_{\mathrm{cl}}(t)= \sqrt{\Lambda_0/3} \left( 1 + 2 \sum_{j=1}^\infty (-1)^j
 \mathrm{e}^{-2 j \alpha \sqrt{\Lambda_0/3}\, t} \right)~.
\label{lateH}
\end{equation}
The classical scale factor for the $\mathrm{sign}(\bar{c}) = -1$ branch in Eq.~\eqref{solhne0} is
\begin{equation}
a_\mathrm{cl}(t) = a_0 \, \cosh^{1/\alpha}\left( \alpha \sqrt{\Lambda_0/3} \, t \right)
 \approx  a_0 \exp\left( \sqrt{\Lambda_0/3} \, t \right)~.
\label{eq_scalelamn0}
\end{equation}
Finally, the energy density is obtained from Eq.~\eqref{eq_Ein_rho} as
\begin{equation}
\rho_\mathrm{cl}(t) =  - (\Lambda_0/8 \pi G_0) \, \cosh^{-2}\left( \alpha \sqrt{\Lambda_0/3} \, t \right)
\approx - (\Lambda_0/2 \pi G_0) \exp\left( -2 \alpha \sqrt{\Lambda_0/3} \, t \right) \approx 0,
\label{eq_energyd_lamn0}.
\end{equation}
Thus, for $\Lambda > 0$, $a_\mathrm{cl}$ at very late times ($t \rightarrow \infty$) grows exponentially
and $\rho_\mathrm{cl}$ effectively vanishes.

\subsection{The identification using the cosmic time}
\label{lambdanon0hub}

We now analyze how the cosmic evolution of the FLRW universe differs in the different branches
of FRG trajectories with $\Lambda_0 > 0$ compared to the branch with $\Lambda_0 = 0$.

The first problem is how to determine the identification of the momentum scale.
If we follow the standard common sense, it is natural to use the Hubble scale. However it becomes constant
in the late time and is not suitable for identification with the momentum scale $k$.
So let us consider the same cutoff scale identification
\bea
k = \frac{\xi}{t},
\eea
which was successful for $\Lambda_0 = 0$.
By inserting this cutoff scale in Eq.~\eqref{eq_GLam}, we can express the running Newton coupling
and cosmological parameter as functions of time as
\begin{eqnarray}
\label{eq_GLam_t_lamne0}
G(t) &=& G_0 \left[ 1 - \tilde\omega G_0 t^{-2} + \tilde\omega_1 G_0^2 t^{-4}
+ \mathcal{O}\left( G_0^3 t^{-6} \right) \right],
\nonumber\\
\Lambda(t) &=& \Lambda_0 \left[ 1 - \tilde\mu G_0 t^{-2} + \tilde\mu_1 G_0^2 t^{-4}
+ \mathcal{O}\left( G_0^3 t^{-6} \right) \right] + t^{-2} \left[ \tilde\nu G_0 t^{-2} + \tilde\nu_1 G_0^2 t^{-4}
+ \mathcal{O}\left( G_0^3 t^{-6} \right) \right],
\end{eqnarray}
where $\tilde\omega \equiv \omega \xi^2, \tilde\omega_1 \equiv \omega_1 \xi^4, \tilde\mu \equiv \mu \xi^2,
\tilde\mu_1 \equiv \mu_1 \xi^4$ and $\tilde\nu \equiv \nu \xi^4, \tilde\nu_1 = \nu_1 \xi^6$.

We can use the Einstein equation \eqref{eq_Ein_H} to obtain the quantum-corrected solution for $H(t)$
as the inverse power series in the cosmic time:
\begin{eqnarray}
H(t) &=& \sqrt{\frac{\Lambda_0}3} \left[ 1 - \frac{\tilde\mu G_0}2 \frac1{t^2}
- \frac{\sqrt3 \tilde\mu G_0}{2 \alpha \sqrt{\Lambda_0}} \frac1{t^3}
+ \left( \frac{\tilde\nu G_0}{2 \Lambda_0} - \frac{\tilde\mu G_0 (18 + \alpha^2 \tilde\mu G_0 \Lambda_0)}
{8 \alpha^2 \Lambda_0} + \frac{\tilde\mu_1 G_0^2}{2} \right) \frac1{t^4} \right.
\nonumber\\
&& \left. + \sqrt{\frac3{\Lambda_0^3}} \left( \frac{\tilde\nu G_0}{\alpha}
- \frac{\tilde\mu G_0 (9 + \alpha^2 \tilde\mu G_0 \Lambda_0)}{2 \alpha^3}
+ \frac{\tilde\mu_1 G_0^2 \Lambda_0}{\alpha} \right) \frac1{t^5} + \mathcal{O}\left( t^{-6} \right) \right],
\end{eqnarray}
which leads to the scale factor
\begin{eqnarray}
a(t) &=& a_0 \, \exp\left[ \sqrt{\frac{\Lambda_0}3} \, t + \sqrt{\frac{\Lambda_0}3} \frac{\tilde\mu G_0}2 t^{-1}
+ \frac{\tilde\mu G_0}{4 \alpha} t^{-2} - \sqrt{\frac3{\Lambda_0}} \left( \frac{\tilde\nu G_0}{18}
- \frac{\tilde\mu G_0 (18 + \alpha^2 \tilde\mu G_0 \Lambda_0)}{72 \alpha^2} + \frac{\tilde\mu_1 G_0^2 \Lambda_0}{18}
\right) t^{-3} + \mathcal{O}\left( t^{-4} \right) \right]
\nonumber\\
&=& a_0 \, \exp\left( \sqrt{\Lambda_0/3} \, t \right) \left[ 1 + \sqrt{\frac{\Lambda_0}3} \frac{\tilde\mu G_0}2 t^{-1}
+ \frac{\tilde\mu G_0}{4} \left( \frac1{\alpha} + \frac{\tilde\mu G_0 \Lambda_0}{6} \right) t^{-2}
+ \mathcal{O}\left( t^{-3} \right) \right].
\end{eqnarray}
By setting the quantum parameters to zero, we should recover the leading-order classical term of the Hubble parameter.
However, the classical solution expands exponentially which can never be reproduced from the power
expansion in $t$.
Another important observation is that, unlike the $\Lambda_0=0$ case, this solution lacks an undetermined constant.
This absence is unusual because the Einstein equation is a first-order differential equation for $H(t)$,
which typically permits a constant of integration. All of these suggest that the cutoff identification
$k = \xi/t$ is not a physically viable choice.


\subsection{The identification using the scale factor}
\label{lambdanon0scale}

In pursuit of a consistent identification of the cutoff scale, we notice the scale factor is another physical quantity
that determines the scale of our universe. So we propose to use it for the identification.
Define
\begin{equation}
\tau = \frac1{2 \alpha \sqrt{\Lambda_0/3}} \exp\left( 2 \alpha \sqrt{\Lambda_0/3} \, t \right)
\propto a_{\rm cl}^{2 \alpha}(t),
\end{equation}
where we have used the cosmological parameter to introduce dimension, and we use this for identification
\bea
k = \frac{\xi}{\tau} \propto a_{\rm cl}^{-2 \alpha}(t),
\label{cf}
\eea
where $\xi$ is a constant of order 1. Here we have put the power $\alpha$ on the scale factor
in analogy with~\eqref{id0}.
The Einstein equation~\eqref{eq_Ein_H} takes the form
\begin{equation}
\sqrt{\Lambda_0/3} \, \tau \frac{dH}{d\tau} = - \frac12 \left( H^2 - \Lambda/3 \right)~.
\label{eqHtau}
\end{equation}
In the literature, this cutoff was employed in~\cite{Bonanno:2001xi} within the consistency condition approach,
where the energy-momentum tensor is conserved along with the right-hand side of the Einstein equation.
In that approach, however, it was shown that no consistent solution exists for a flat FLRW universe,
except in the presence of exotic matter.

Here we demonstrate that this cutoff choice is a viable option for the case $\Lambda_0 > 0$ if
we use the modified continuity equation, in contrast to the consistency condition approach.
It is also important to note that we select $\Lambda_0$ to match the dimension of the momentum of the cutoff scale
$k$, unlike the $\Lambda_0 = 0$ case, where $G_0$ is chosen. This is because the quantum effects in the solution
is expected to become strong around the order of Planck scale: If we had used $G_0$ with positive power
(for dimensional reason) in $\tau$, the Planck scale in the quantum effects would cancel out
from Eq.~\eqref{eq_GLam}, in contradiction to this expectation.

With the choice of cutoff~\eqref{cf}, the scale-dependent cosmological parameter~\eqref{eq_GLam} becomes
\begin{equation}
\Lambda(\tau) = \Lambda_0 \left[ 1 - \tilde\mu G_0 \tau^{-2} + \tilde\mu_1 G_0^2 \tau^{-4}
+ \mathcal{O}\left( G_0^3 \tau^{-6} \right) \right] + \tau^{-2} \left[ \tilde\nu G_0 \tau^{-2}
+ \tilde\nu_1 G_0^2 \tau^{-4} + \mathcal{O}\left( G_0^3 \tau^{-6} \right) \right]~.
\label{lamtau}
\end{equation}
By substituting this into Eq.~\eqref{eqHtau}, we obtain the solution
for the Hubble parameter in terms of $\tau$ as
\begin{equation}
H(\tau) = \sqrt{\Lambda_0/3} \left[ 1 + \frac{2 c'}{\tau} + \frac{2 c'^2}{\tau^2} + \frac{2 c'^3}{\tau^3}
 + \frac{2 c'^4}{\tau^4}
+ \frac{\tilde\mu G_0}{2 \tau^2} + \frac{\tilde\mu G_0 c'}{2 \tau^3}
+ \frac{2 \tilde\mu G_0 c'^2}{3 \tau^4} + \frac{(\tilde\mu^2 - 4 \tilde\mu_1) G_0^2}{24 \tau^4}
- \frac{\tilde\nu G_0}{6 \Lambda_0 \tau^4} + \mathcal{O}\left( \tau^{-5} \right) \right]~.
\label{solHtau}
\end{equation}
As discussed before for $\Lambda_0=0$, the time translation symmetry is broken here due to the time
dependence of the cosmological term in the Einstein equation. Consequently, the new parameter $c'$ plays
a significant role in obtaining the quantum-corrected solution for the Hubble parameter.
However, as mentioned for the $\Lambda_0 = 0$ case,
to recover the classical solution, we should set the quantum parameters $\tilde\mu$, $\tilde\nu$ and
$\tilde\mu_1$ to zero. This allows us to absorb the free parameter within time or set it to zero.
This point will become clearer in the following steps.

If we set $\tilde\mu$, $\tilde\nu$ and $\tilde\mu_1$  to zero, the expansion of the Hubble parameter~\eqref{solHtau}
becomes
\begin{eqnarray}
H_\mathrm{cl} & = &\sqrt{\frac{\Lambda_0}{3}}\left[1 + \frac{2 c'}{\tau} + \frac{2 c'^2}{\tau^2}
+ \frac{2 c'^3}{\tau^3} + \frac{2 c'^4}{\tau^4} + \cdots \right]  \nonumber \\
& = & \sqrt{\frac{\Lambda_0}{3}}\left( 1 + \frac{c'}{\tau} \right) \left( 1 + \frac{c'}{\tau}
+ \frac{c'^2}{\tau^2} + \frac{c'^3}{\tau^3} + \frac{c'^4}{\tau^4} + \cdots \right)
=\sqrt{\frac{\Lambda_0}{3}} \frac{1 + c'/\tau}{1 - c'/\tau}~.
\label{clHtau}
\end{eqnarray}
Redefining a new free parameter $c$
\begin{equation}
c = \frac{\ln\left( 2 \alpha \sqrt{\Lambda_0/3} \, |c'| \right)}{2 \alpha \sqrt{\Lambda_0/3}}
\quad \Rightarrow \quad
2 \alpha \sqrt{\Lambda_0/3} \, c' = \mathrm{sign}(c') \, \mathrm{e}^{2 \alpha \sqrt{\Lambda_0/3} \, c},
\label{c'relc}
\end{equation}
we obtain the exact solution for the Hubble parameter:
\begin{equation}
H_\mathrm{cl} = \sqrt{\frac{\Lambda_0}{3}}
\frac{1 + \mathrm{sign}(c')\mathrm{e}^{-2 \alpha \sqrt{\Lambda_0/3} (t - c)}}{1 - \mathrm{sign}(c')
\mathrm{e}^{-2 \alpha \sqrt{\Lambda_0/3} (t - c)}}
= \begin{cases} \sqrt{\frac{\Lambda_0}{3}} \tanh\left( \alpha \sqrt{\Lambda_0/3} (t - c) \right),
& \mathrm{sign}(c') = -1, \\ \sqrt{\frac{\Lambda_0}{3}} \coth\left( \alpha \sqrt{\Lambda_0/3} (t - c) \right),
& \mathrm{sign}(c') = 1. \end{cases}
\end{equation}
Here the parameter $c$ can be either absorbed into the cosmic time or set to $c = 0$. However,
in the case of the quantum-corrected solution, the free parameter $c$, or more precisely $|c'|$,
acquires a physical meaning due to quantum corrections through $\Lambda(t)$. This becomes evident
when re-expressing the quantum-corrected solution as
\begin{equation}
H(\tau) = \sqrt{\Lambda_0/3} \left[ 1 + \frac{2 |c'|}{\tau} \left( \frac{\tau_c}{\tau} + \mathrm{sign}(c') \right)
+ \frac{2 |c'|^2}{\tau^2} \left( \frac{\tau_c}{\tau}\mathrm{sign}(c') +1 \right) + \cdots \right]~,
\label{qHlam0ne0}
\end{equation}
where the parameter $|c'|$ gives a critical scale determined by
\begin{equation}
\tau_c = \tilde\mu G_0 / 4 |c'|~.
\end{equation}
If $c' > 0$, which implies $\mathrm{sign}(c') = 1$, the quantum correction enhances the Hubble parameter eternally.
On the other hand, if $c' < 0$, meaning $\mathrm{sign}(c') = -1$, the infintely large late-time expansion rate
remains constant at $\sqrt{\Lambda_0/3}$ as the quantum corrections become negligible. However, the value of
the constant is still affected by the quantum correction terms at late time. Since the leading-order quantum
correction is stronger than the subleading orders, we observe from \eqref{qHlam0ne0} that late-time cosmic
expansion is suppressed due to quantum corrections compared to the very late-time classical behavior
when $\tau > \tau_c$, and the opposite occurs for $\tau < \tau_c$.

In reality, based on Eq.~\eqref{c'relc}, $|c'|$ is found to be very large since the leading-order parameter $c$
is of the order of the present time, making $\tau_c$ unreasonably small. As a result, the $\tau > \tau_c$ case
dominates. While the quantum corrections are small, they produce an effect opposite to what is expected from
classical behavior. Specifically, while the present universe is accelerating, the quantum-corrected Hubble parameter
appears to be smaller than the classical Hubble parameter. Even in the case of $\Lambda_0 = 0$,
the quantum-corrected Hubble parameter increases, contrary to the classical expectation of a decelerating universe.

Since the Hubble parameter is defined in terms of $\tau$ as
$H(\tau) = a^{-1} (da/d\tau) (d\tau/dt)$, the scale factor can be expressed as
\begin{equation}
a(\tau) = \tau^{1/2 \alpha} \left[ 1 - \frac{c'}{\alpha \tau}- \left( \frac{4 c'^2 + \tilde{\mu} G_0}{8 \alpha}
 - \frac{c'^2}{2 \alpha^2} \right) \frac{1}{\tau^2} + \mathcal{O}\left( \tau^{-3} \right) \right]~.
\end{equation}
The energy density can be computed straightforwardly from Eq.~\eqref{eq_Ein_rho} as
\begin{equation}
\rho(\tau) = \frac{\Lambda_0 c'}{2 \pi G_0 \tau} + \frac{\Lambda_0 (4 c'^2 + \tilde\mu G_0)}{4 \pi G_0 \tau^2}
+ \frac{3 \Lambda_0 c' (4 c'^2 + \tilde\mu G_0 + 4 \tilde\omega G_0/3)}{8 \pi G_0 \tau^3}
+ \mathcal{O}\left( \tau^{-4} \right).
\end{equation}
As expected, the entire contribution to the energy density at late times arises from quantum corrections.
For $c' < 0$, the critical scale $\tau_c$ emerges from the competition between the first term and the term
$\Lambda_0 \tilde\mu G_0/(4 \pi G_0 \tau^2)$ which appears in the parenthesis of the second term.

It is worth noting that the quantum-corrected solution for the Hubble parameter under this cutoff choice
addresses the issue of requiring an additional integration constant. Furthermore, this solution reproduces
exponential subleading terms in $H(t)$ for late times, in contrast to the power expansion terms obtained with
the $k = \xi/t$ choice. Therefore, it can be concluded from the quantum-corrected solution
for the Hubble parameter that the late-time behaviour aligns with the classical solution,
with minor quantum improvements. This demonstrates the viability of the proposed cutoff choice,
unlike the earlier $k = \xi/t$ choice.

\section{Summary and Conclusions} \label{conclu}

In this work, we have studied a spatially flat FLRW universe in late time, taking quantum gravitational effects
into account, within the framework of asymptotically safe gravity. The FRG flow of the effective average action
in asymptotically safe gravity has resulted in the running of the Newton coupling and the cosmological
parameter under the Einstein-Hilbert truncation. We have analyzed the cosmic evolution of the FLRW cosmology
at late times by improving the Einstein equations to include the scale dependence of Newton coupling and
the cosmological parameter. Specifically, we have considered the conservation of the entire right-hand side
of the Einstein equations, which has led to a modified continuity equation. The main focus of this work has
been to identify a viable cutoff scale, $k$, that can yield a consistent FLRW cosmology.

In this context, we have examined cosmological solutions for two branches of FRG trajectories.
The first branch has involved a power series expansion of the dimensionful cosmological parameter, $\Lambda(t)$,
beginning with a $k^4$ term, with $\Lambda_0=0$. This trajectory has approached the Gaussian fixed point
as $k \rightarrow 0$ and has been referred to in the literature as
the ``separatrix"~\cite{Reuter:2001ag, Codello:2008vh}. The second branch, which may be realized in nature,
starts with a $k^2$ term in the expansion of the cosmological parameter and is valid up to a certain
non-zero momentum scale, where the $\beta$-function becomes singular as the dimensionless cosmological
parameter $\lambda \equiv \Lambda(k)/k^2$ comes to a ``singular line'' near $1/2$.
(The beta function becomes singular as can be seen from Eq.~\eqref{terms}. Similar singularity exists for other
choices of the regulators. The RG flow near this singularity is discussed in detail in Ref.~\cite{KO}.)

For studying cosmology in these two branches, we have considered the cutoff scale $k$ in terms of either classical
or quantum Hubble parameter for the $\Lambda_0 = 0$ case. We have chosen $k \sim H_{\rm cl}(t)$, which measures
the curvature of the FLRW spacetime. Before incorporating quantum corrections through the Einstein equations,
there was the time translation symmetry, implying that the integration constant in the classical solution
does not carry any physical significance. However, in the quantum-corrected solution, a new constant comes
into the solution with physical significance through the critical time scale $t_c$ which defines two distinct
phases of the quantum-corrected Hubble parameter: the repulsive phase (dS phase) and the attractive phase (AdS phase).
In the attractive phase, the leading quantum correction has introduced an attractive effect, causing the expansion
rate of the FLRW universe to slow down compared to the classical expansion rate,
$H(t) < H_{\mathrm{cl}}(t) = 1/(\alpha t)$ for $t < t_{c}$. In the repulsive phase (dS phase),
it has enhanced the expansion rate beyond the classical rate when $t > t_{c}$.

Considering another possibility that the cutoff scale $k$ may have a simple functional relationship with
the quantum Hubble parameter rather than with the classical Hubble parameter, we examined the cutoff choice
$k(H) = \xi G_0^{(\beta - 1)/2} H^\beta$, where $\beta$ is a constant. In earlier work~\cite{Mandal:2020umo},
three specific values of $\beta$ namely, $\beta = 1/4, 3/4$ and 1 were analyzed, and the corresponding entropy
generation due to quantum effects was investigated. It was observed that entropy generation diverges at late times
for $\beta = 1/4$, indicating that this cutoff choice is not suitable for studying cosmology.
On the other hand, for $\beta = 3/4$ and $\beta = 1$, entropy generation converges to a constant and thus these cutoff choices are viable. This suggests the existence of a critical value of $\beta$ that
determines the appropriateness of a cutoff choice. By studying the asymptotic behaviour of the Hubble parameter
at large times, we have demonstrated that the critical value is $\beta = 1/2$. This value marks
a transition in the scale factor's behaviour from power-law growth ($\beta \geq 1/2$) to exponential growth
($\beta < 1/2$). From the exponential behaviour of the scale factor for $\beta < 1/2$, we infer that
such cutoff choices are not viable, as the large momentum scale $k$ introduces significant perturbations
that entirely disrupt the classical power-law behaviour for the $\Lambda_0=0$ trajectory.
On the other hand, for $\beta \geq 1/2$, the behaviour follows a power law similar to the classical case.
Here the perturbations remain small enough to preserve this expected behaviour with minor quantum corrections.
Therefore, the cutoff identification is valid for $\beta \geq 1/2$ making it suitable for cosmological studies.

For trajectories with $\Lambda_0 > 0$, the Hubble parameter is not suitable for the identification because
it becomes constant quickly. Following the successful case for $\Lambda_0=0$, we tried to use the cutoff scale
$k = \xi/t$ in terms of the cosmic time, but found that it does not provide a consistent solution because
the quantum corrections to the Hubble parameter introduce power expansion terms that become more
significant than the classical subleading exponential terms and cannot reproduce the classical solution
in the classical limit.
Moreover, unlike the $\Lambda_0=0$ case, this solution with this cutoff lacks an undetermined constant term,
despite the fact that the Einstein equation is a first-order differential equation for $H(t)$.
All of these suggest that the cutoff identification $k = \xi/t$ may not represent a physically viable choice.
In contrast, we can resolve this issue with  $k = \xi/\tau$ which indicates that the quantum solution
resembles the classical solution, along with quantum improvements in the same form. We expect
that the perturbations introduced by quantum corrections should not be too large to significantly alter
the form of the solutions at late times. Through this analysis, we have demonstrated a method to evaluate
the viability of different cutoff scale choices for two branches of the trajectories $\Lambda_0 = 0$ and
$\Lambda_0 > 0$.

\acknowledgments

We would like to thank Akihiro Ishibashi for valuable discussions at the early stage of this work.
The work of C.M.C. was supported by the National Science and Technology Council of the R.O.C. (Taiwan) under the grant NSTC 113-2112-M-008-027.
The work of R.M. was supported by the National Science and Technology Council of the R.O.C. (Taiwan) under the grant NSTC 113-2811-M-008-046.
The work of N.O. was supported in part by the Grant-in-Aid for Scientific Research Fund of the JSPS (C) No. 20K03980.

\begin{appendix}

\section{Perturbative Solution of Running Couplings} \label{sec_SolRG}

The FRG equations for running couplings $G(k)$ and $\Lambda(k)$ for Einstein-Hilbert truncation
are~\cite{Reuter:1996cp}
\begin{eqnarray}
k \partial_k G(k) &=& \eta_N G(k),
\\
k \partial_k \Lambda(k) &=& \eta_N \Lambda(k) + \frac{k^4 G(k)}{2 \pi}
\left[ 10 \Phi^1_2(-2 \lambda) - 8 \Phi^1_2(0) - 5 \eta_N \tilde\Phi^1_2(-2 \lambda) \right],
\end{eqnarray}
where $\eta_N$ is the anomalous dimension of $\sqrt{g} R$:
\begin{equation}
\eta_N = \frac{k^2 G(k) \, B_1(\lambda)}{1 - k^2 G(k) \, B_2(\lambda)},
\end{equation}
in which the dimensionless coupling is defined as $\lambda = \Lambda(k)/k^2$, and
$B_1(\lambda)$ and $B_2(\lambda)$ are abbreviations of
\begin{equation}
B_1(\lambda) = \frac1{3 \pi} \left[ 5 \Phi^1_1(-2 \lambda) - 18 \Phi^2_2(-2 \lambda) - 4 \Phi^1_1(0)
 - 6 \Phi^2_2(0) \right], \quad
B_2(\lambda) = - \frac1{6 \pi} \left[ 5 \tilde\Phi^1_1(-2 \lambda) - 18 \tilde\Phi^2_2(-2 \lambda) \right].
\end{equation}
The threshold functions that appear here are defined as
\begin{equation}
\Phi^p_n(w) = \frac1{\Gamma(n)} \int_0^\infty dz z^{n - 1} \frac{R^{(0)}(z)
 - z \partial_z R^{(0)}(z)}{\left[ z + R^{(0)}(z) + w \right]^p}, \qquad
{\tilde\Phi}^p_n(w) = \frac1{\Gamma(n)} \int_0^\infty dz z^{n - 1} \frac{R^{(0)}(z)}{\left[ z + R^{(0)}(z)
 + w \right]^p},
\end{equation}
which depend on the choice the cutoff function $R^{(0)}(z)$. Throughout this paper, $\Gamma(n)$ is the gamma function.

The FRG equations can be further reexpressed as
\begin{eqnarray}
\label{FRG1}
k \partial_k G(k) &=& \frac{k^2 G^2(k) \, B_1(\lambda)}{1 - k^2 G(k) \, B_2(\lambda)},
\\
k \partial_k \Lambda(k) &=& - \frac{k^2 G(k) \left\{ k^4 G(k) \left[ A_1(\lambda) \, B_2(\lambda)
 + A_2(\lambda) \, B_1(\lambda) \right] - A_1(\lambda) \, k^2 - B_1(\lambda) \, \Lambda(k) \right\}}{1 - k^2 G(k)
 \, B_2(\lambda)},
\label{FRG2}
\end{eqnarray}
with two additional abbreviations
\begin{equation}
A_1(\lambda) = \frac1{\pi} \left[ 5 \Phi^1_2(-2 \lambda) - 4 \Phi^1_2(0) \right], \qquad
A_2(\lambda) = \frac{5}{2 \pi} {\tilde\Phi}^1_2(-2 \lambda).
\end{equation}

For small value of $k$, appropriate to late time behavior, the running couplings are assumed as
\begin{equation}
\label{eq_GL-expand}
G(k) = \sum_{j=0} G_j k^j, \qquad \Lambda(k) = \sum_{j=0} \Lambda_j k^j.
\end{equation}
In what follows, we solve the FRG equations perturbatively.

\subsection{Exponential Cutoff}

In~\cite{Bonanno:2001xi}, the low energy expansions of running couplings were derived with the exponential cutoff
\begin{equation}
R^{(0)}(z) = \frac{z}{\exp(z) - 1}.
\end{equation}

\noindent{\bf Case I: $\Lambda_0=0$}

In order to solve the FRG equations perturbatively, one should expand the factor
$(Z - 2 \Lambda(k)/k^2)^{-p}$ with $Z = z + R^{(0)}$ in power of $k$. For $\Lambda_0 = 0$, we write
\begin{equation}
\left( Z - \frac{2 \Lambda(k)}{k^2} \right)^{-p} = Z^{-p} \left[ 1 + \frac{2 p \Lambda_4}{Z} k^2
+ \left( \frac{2 p \Lambda_6}{Z} + \frac{2 p (p + 1) \Lambda_4^2}{Z^2} \right) k^4 + \cdots \right],
\end{equation}
and then the threshold functions $\Phi^p_n(-2 \lambda)$ can be expanded as
\begin{equation}
\Phi^p_n\left( - \frac{2 \Lambda(k)}{k^2} \right) = \Phi^p_n(0) + 2 p \Lambda_4 \Phi^{p+1}_n(0) k^2
+ \left[ 2 p \Lambda_6 \Phi^{p+1}_n(0) + 2 p (p + 1) \Lambda_4^2 \Phi^{p+2}_n(0) \right] k^4 + \cdots,
\end{equation}
and similarly for $\tilde\Phi^p_n(-2 \lambda)$ as
\begin{equation}
\tilde\Phi^p_n\left( - \frac{2 \Lambda(k)}{k^2} \right)
= \tilde\Phi^p_n(0) + 2 p \Lambda_4 \tilde\Phi^{p+1}_n(0) k^2 + \left[ 2 p \Lambda_6 \tilde\Phi^{p+1}_n(0)
+ 2 p (p + 1) \Lambda_4^2 \tilde\Phi^{p+2}_n(0) \right] k^4 + \cdots.
\end{equation}
By solving the FRG equation order by order, we determine the leading terms of the solutions~\eqref{eq_GL-expand} as
\begin{eqnarray}
&& G_2 = \frac{\Phi^1_1(0) - 24 \Phi^2_2(0)}{6 \pi} G_0^2,
\nonumber\\
&& G_4 = \frac{2 \left[ \Phi^1_1(0) - 24 \Phi^2_2(0) \right]^2 - \left[ \Phi^1_1(0)
 - 24 \Phi^2_2(0) \right] \left[ 5 {\tilde\Phi}^1_1(0) - 18 {\tilde\Phi}^2_2(0) \right]
 + 3 \Phi^1_2(0) \left[ 5 \Phi^2_1(0) - 36 \Phi^3_2(0) \right]}{72 \pi^2} G_0^3,
\nonumber\\
&& \Lambda_4 = \frac{\Phi^1_2(0)}{4 \pi} G_0, \qquad \Lambda_6 = \frac{\left[ \Phi^1_1(0)
 - 24 \Phi^2_2(0) \right] \left[ 3 \Phi^1_1(0) - 10 {\tilde\Phi}^1_2(0) \right]
 + 30 \Phi^1_2(0) \Phi^2_2(0)}{72 \pi^2} G_0^2,
\end{eqnarray}
with the specific values
\begin{eqnarray} \label{eq_Phi_val}
&& \Phi^0_1(0) = \frac{\pi^2}{3}, \qquad
\Phi^1_1(0) = \frac{\pi^2}{6}, \qquad
\Phi^1_2(0) = 2 \zeta(3), \qquad
\Phi^2_1(0) = 1, \qquad \Phi^2_2(0) = 1, \qquad
\Phi^3_2(0) = \frac12,
\nonumber\\
&& {\tilde\Phi}^1_1(0) = 1, \qquad
{\tilde\Phi}^1_2(0) = 1, \qquad
{\tilde\Phi}^2_1(0) = \ln 2, \qquad
{\tilde\Phi}^2_2(0) = \frac12,
\end{eqnarray}
where $\zeta(3)$ is the value of the zeta function $\zeta(n)$ at $n=3$.
The solutions agree with the results in~\cite{Bonanno:2001xi} as, in the notation of~\eqref{eq_GLam}: we find  $\Lambda_2 =0$, $\Lambda_n = G_n = 0$ for odd $n$ and $G_2 = - \omega G_0^2, G_4 = \omega_1 G_0^3, \Lambda_4 = \nu G_0, \Lambda_6 = \nu_1 G_0^2$,
\begin{equation}
\label{eq_exp_0}
\omega = \frac{4}{\pi} \left( 1 - \frac{\pi^2}{144} \right), \quad
\nu = \frac{\zeta(3)}{2 \pi}, \quad
\omega_1 = \omega^2 - \frac{\omega}{3 \pi} - \frac{13 \nu}{6 \pi}, \quad
\nu_1 = - \omega \nu + \frac{5 \omega}{6 \pi} + \frac{5 \nu}{3 \pi}.
\end{equation}

\noindent{\bf Case II: $\Lambda_0 \ne 0$}

The expansion of the factor $(Z - 2 \Lambda(k)/k^2)^{-p}$ for $\Lambda_0 \ne 0$ is written as
{\small \begin{eqnarray}
\left( Z - \frac{2 \Lambda(k)}{k^2} \right)^{-p} &=& \frac{(-1)^p}{2^p \Lambda_0^p} k^{2p}
- \frac{(-1)^p p \Lambda_1}{2^p \Lambda_0^{p+1}} k^{2p+1}
+ \frac{(-1)^p}{2^{p+1}} \left( \frac{p (p + 1) \Lambda_1^2 - 2 p \Lambda_0 \Lambda_2}{\Lambda_0^{p+2}}
+ \frac{p Z}{\Lambda_0^{p+1}} \right) k^{2p+2}
\nonumber\\
&+& \frac{(-1)^p}{2^{p+1}} \left(  \frac{p (p + 1) (p + 2) \Lambda_1^3 + 6 p \Lambda_0 \Lambda_3
- 6 p (p + 1) \Lambda_0 \Lambda_1 \Lambda_2}{3 \Lambda_0^{p+3}}
+ \frac{p (p+1) \Lambda_1 Z}{\Lambda_0^{p+2}} \right) k^{2p+3} + \cdots,
\end{eqnarray} }
and then it leads to the expansion of $\Phi^p_n(-2 \lambda)$
{\small \begin{eqnarray}
\Phi^p_n\left( - \frac{2 \Lambda(k)}{k^2} \right) &=& \frac{(-1)^p}{2^p \Lambda_0^p} \Phi^0_n(0) k^{2p}
- \frac{(-1)^p p \Lambda_1}{2^p \Lambda_0^{p+1}} \Phi^0_n(0) k^{2p+1}
+ \frac{(-1)^p}{2^{p+1}} \left( \frac{p (p + 1) \Lambda_1^2 - 2 p \Lambda_0 \Lambda_2}{\Lambda_0^{p+2}} \Phi^0_n(0)
+ \frac{p}{\Lambda_0^{p+1}} \Phi^{-1}_n(0) \right) k^{2p+2}
\nonumber\\
&+& \frac{(-1)^p}{2^{p+1}} \left(  \frac{p (p + 1) (p + 2) \Lambda_1^3 + 6 p \Lambda_0^2 \Lambda_3
- 6 p (p + 1) \Lambda_0 \Lambda_1 \Lambda_2}{3 \Lambda_0^{p+3}} \Phi^0_n(0)
+ \frac{p (p+1) \Lambda_1}{\Lambda_0^{p+2}} \Phi^{-1}_n(0) \right) k^{2p+3} + \cdots,
\end{eqnarray} }
and similarly for ${\tilde\Phi}^p_n(-2 \lambda)$. The FRG equations lead to
$0 = G_1 = G_3 = \cdots, 0 = \Lambda_1 = \Lambda_3 = \cdots$ and
\begin{eqnarray}
	&& G_2 = - \frac{2 \Phi^1_1(0) + 3 \Phi^2_2(0)}{3 \pi} G_0^2, \qquad
	G_4 = \frac{\left[ 2 \Phi^1_1(0) + 3 \Phi^2_2(0) \right]^2}{9 \pi^2} G_0^3
	- \frac{5 \Phi^0_1(0)}{24 \pi} \frac{G_0^2}{\Lambda_0},
	\nonumber\\
	&& \Lambda_2 = - \frac{2 \Phi^1_1(0) + 3 \Phi^2_2(0)}{3 \pi} G_0 \Lambda_0, \qquad
	\Lambda_4 = \frac{\left[ 2 \Phi^1_1(0) + 3 \Phi^2_2(0) \right]^2}{9 \pi^2} G_0^2 \Lambda_0
	- \frac{5 \Phi^0_1(0) + 24 \Phi^1_2(0)}{24 \pi} G_0,
	\\
	&& \Lambda_6 = - \frac{\left[ 2 \Phi^1_1(0) + 3 \Phi^2_2(0) \right]^3 G_0^3 \Lambda_0}{27 \pi^3}
		+ \frac{\left[ 2 \Phi^1_1(0) + 3 \Phi^2_2(0) \right] \left[ 10 \Phi^0_1(0) + 36 \Phi^1_2(0) - 5 \tilde\Phi^0_1(0) \right] G_0^2}{108 \pi^2} - \frac{\left[ 48 \Phi^0_2(0) + 5 \Phi^{-1}_1(0) \right] G_0}{72 \pi \Lambda_0}. \nonumber
\end{eqnarray}
With the specific values~\eqref{eq_Phi_val} and
\begin{equation}
	 \Phi^{-1}_1(0) = \frac{\pi^2}2 + 3 \zeta(3), \qquad \Phi^0_2(0) = 6 \zeta(3), \qquad \tilde\Phi^0_1(0) = \frac{\pi^2}6, 
\end{equation}
we have
\begin{eqnarray}
	G(k) &=& G_0 \left[ 1 - \frac{9 + \pi^2}{9 \pi} G_0 k^2 + \left[ \left( \frac{9 + \pi^2}{9 \pi} \right)^2
	- \frac{5 \pi}{72} \frac1{G_0 \Lambda_0} \right] G_0^2 k^4 + \mathcal{O}(G_0^3k^6) \right],
	\\
	\Lambda(k) &=& \Lambda_0 \left[ 1 - \frac{9 + \pi^2}{9 \pi} G_0 k^2
	+ \left( \frac{9 + \pi^2}{9 \pi} \right)^2 G_0^2 k^4 - \left(\left( \frac{9 + \pi^2}{9 \pi} \right)^3 + \frac{5 \pi^2 + 606 \zeta(3)}{144 \pi \Lambda_0^2 G_0^2}\right)G_0^3k^6  +\mathcal{O}(G_0^4 k^8) \right]
 \nn
&&  	+ G_0 k^4 \left[  - \frac{5 \pi^2 + 144 \zeta(3)}{72 \pi} +   \frac{ (9 + \pi^2)(5 \pi^2 + 144 \zeta(3))}{648 \pi^2}
		 G_0 k^2  + \mathcal{O}(G_0^2 k^4) \right].
\end{eqnarray}
This result gives the coefficients in Eq.~\eqref{eq_GLam} for the exponential cutoff.
 Note that these two cases are ``disconnected'', i.e. the coefficients in the case $\Lambda_0 = 0$
	are not the $\Lambda_0 \to 0$ limit of the related coefficients in the case $\Lambda_0 \ne 0$.

\subsection{Optimized Cutoff}

For the optimized cutoff~\cite{Codello:2008vh}, the cutoff function is chosen as
\begin{equation}
R^{(0)}(z) = (1 - z) \Theta(1 - z),
\end{equation}
then the integration of threshold functions can be carried out
\begin{equation}
\Phi^p_n(w) = \frac{(1 + w)^{-p}}{\Gamma(n + 1)}, \qquad \tilde\Phi^p_n(w) = \frac{(1 + w)^{-p}}{\Gamma(n + 2)}.
\end{equation}
The four abbreviations in FRG equations~\eqref{FRG1} and \eqref{FRG2} can be simplified
\begin{equation}
A_1 = \frac{k^2 + 8 \Lambda}{2 \pi (k^2 - 2 \Lambda)}, \quad
A_2 = \frac{5 k^2}{12 \pi (k^2 - 2 \Lambda)}, \quad
B_1 = - \frac{11 k^4 - 18 k^2 \Lambda + 28 \Lambda^2}{3 \pi (k^2 - 2 \Lambda)^2}, \quad
B_2 = \frac{k^4 + 10 k^2 \Lambda}{12 \pi (k^2 - 2 \Lambda)^2}.
\label{terms}
\end{equation}

\noindent{\bf Case I: $\Lambda_0 = 0$}

For the case $\Lambda_0 = 0$, the leading orders of equations lead to $\Lambda_1 = \Lambda_2 = \Lambda_3 = 0$
and the solutions are
\begin{equation} \label{eq_Optimized_0}
G(k) = G_0 \left[ 1 - \frac{11}{6 \pi} G_0 k^2 + \frac{217}{72 \pi^2} G_0^2 k^4
+ \mathcal{O}\left( G_0^3 k^6 \right) \right], \qquad \Lambda(k)
= G_0 k^4 \left[ \frac1{8 \pi} + \frac7{54 \pi^2} G_0 k^2 + \mathcal{O}\left( G_0^2 k^4 \right) \right].
\end{equation}

\noindent{\bf Case II: $\Lambda_0 \ne 0$}

The solutions for the case $\Lambda_0 \ne 0$ are
\begin{eqnarray}
G(k) &=& G_0 \left[ 1 - \frac{7}{6 \pi} G_0 k^2 + \left( \frac{49}{36 \pi^2}
- \frac{5}{24 \pi G_0 \Lambda_0} \right) G_0^2 k^4  + \mathcal{O}\left( G_0^3 k^6 \right) \right],
\nonumber\\
\Lambda(k) &=& \Lambda_0 \left[ 1 - \frac{7}{6 \pi} G_0 k^2 + \frac{49}{36 \pi^2} G_0^2 k^4
- \left( \frac{29}{72 \pi G_0^2 \Lambda_0^2} + \frac{343}{216 \pi^3} \right) G_0^3 k^6
+ \mathcal{O}\left( G_0^4 k^8 \right) \right]
\nonumber\\
&& + G_0 k^4 \left[ - \frac{17}{24 \pi} + \frac{119}{144 \pi^2} G_0 k^2 + \mathcal{O}\left( G_0^2 k^4 \right) \right].
\end{eqnarray}

 Again these two cases are ``disconnected'' as in the exponential cutoff.

\end{appendix}

\end{document}